\documentclass[11pt]{article}
\usepackage{graphicx}
\usepackage{multirow}

\newcommand{\BABARPubYear}    {10}

\newcommand{\BABARProcNumber} {098}
\newcommand{\SLACPubNumber} {14309}

\usepackage{fullpage}

\usepackage{maybemath}
\usepackage[italic]{hepparticles}
\usepackage{heppennames}

\DeclareRobustCommand{\PsX}{\HepParticle{X}{\Pqs}{}\xspace}
\DeclareRobustCommand{\PcX}{\HepParticle{X}{\Pqc}{}\xspace}
\DeclareRobustCommand{\PcsX}{\HepParticle{X}{\Pqc(\Pqs)}{}\xspace}
\DeclareRobustCommand{\PuX}{\HepParticle{X}{\Pqu}{}\xspace}

\usepackage{relsize}
\def\babar{\mbox{\slshape B\kern-0.1em{\smaller A}\kern-0.1em
    B\kern-0.1em{\smaller A\kern-0.2em R}}}

\def\bxclnu{\HepProcess{\PB\to\PcX\Pl\Pgn}{}\xspace}
\def\bxulnu{\HepProcess{\PB\to\PuX\Pl\Pgn}{}\xspace}

\def\bxsg{\ensuremath{\PB \to X_{s} \Pgg}\xspace}
\def\bxdg{\ensuremath{\PB \to X_{d} \Pgg}\xspace}
\def\bxsdg{\ensuremath{\PB \to X_{s+d} \Pgg}\xspace}
\def\bkg        {\ensuremath{\PB \to \PKst \Pgg}\xspace}

\def\CP         {\ensuremath{C\!P}\xspace}
\def\mb         {\ensuremath{m_{b} }\xspace}  
\def\acp        {\ensuremath{A_{CP}}\xspace}
\newcommand{\gev}{\ensuremath{\mathrm{\,Ge\kern -0.1em V}}\xspace}
\newcommand{\mev}{\ensuremath{\mathrm{\,Me\kern -0.1em V}}\xspace}
\def\BB         {\ensuremath{\PB{}\PaB}\xspace}
\def\egcms      {\ensuremath{E^{*}_{\gamma}}\xspace}
\def\pizeta     {\Pgpz{}(\Pgh)\xspace}
\newcommand{\vs}{\mbox{\textsl{vs.}}\xspace}

\begin{document}
{\pagestyle{empty}

\begin{flushright}
SLAC-PUB-\SLACPubNumber \\
\babar-PROC-\BABARPubYear/\BABARProcNumber \\
November 2010 \\
\end{flushright}

\par\vskip 4cm

% Title of the paper
\begin{center}
\Large \bf Inclusive Measurements of \bxclnu and \bxsg Decays: Mini-review
\end{center}
\bigskip

\begin{center}
\large 
Kyle J. Knoepfel\\
University of Notre Dame\footnote{now at Fermi National Accelerator Laboratory} \\
E-mail: {\tt knoepfel@fnal.gov} \\
(from the \babar\ Collaboration)
\end{center}
\bigskip \bigskip

% Abstract
\begin{center}
\large \bf Abstract
\end{center}
A mini-review of measurements of inclusive semileptonic \bxclnu and
radiative \bxsg decays is presented.  The semileptonic \PcX mass
moments are presented from Belle and \babar.  The inclusive \bxsg
branching fraction is presented from Belle as well as a preliminary
measurement of the direct \CP-asymmetry of \bxsdg decays at \babar.
Fundamental Standard Model parameters and heavy quark parameters can
be derived from the \PcX mass and lepton energy moments from \bxclnu
decays and from the photon energy moments from \bxsg decays.  The
values of the CKM matrix element $|V_{cb}|$, the \Pqb-quark mass \mb,
and the Fermi motion of the \Pqb-quark inside the \PB-meson are
presented based on a global fit to these moments by the Heavy Flavor
Averaging Group.

\vfill
\begin{center}
Contributed to the Proceedings of the X$^{th}$ Nicola Cabibbo
International Conference on Heavy Quarks and Leptons, \\
October 11--15, 2010, Frascati (Rome), Italy
\end{center}

%\vspace{1.0cm}
%\begin{center}
%{\em Stanford Linear Accelerator Center, Stanford University, 
%Stanford, CA 94309} \\ \vspace{0.1cm}\hrule\vspace{0.1cm}
%Work supported in part by Department of Energy contract DE-AC03-76SF00515.
%\end{center}
%
}

\section{Introduction}

Inclusive semileptonic \bxclnu and radiative \bxsg decays, where \PcsX
represents any final hadronic state with unit charm (strangeness), are
powerful laboratories for new and Standard Model (SM) physics.  The
optical theorem can be used to related the inclusive decay rates to
the forward-scattering of the \PB-meson.  The resulting expression is
the basis for an operator product expansion (OPE) in powers of
$\Lambda/m_B$, where $\Lambda$ is the scale of the momentum transfer
of the decay, and $m_B$ is the \PB-meson mass.

Since the lowest-order non-perturbative term arises at roughly
$1/m_B^2$, calculations of inclusive measurements are typically
precise (less than $10\%$ uncertainty) as the theoretical predictions
are free from large uncertainties that can arise from hadronic form
factors present in exclusive decays.  The precision on inclusive
calculations of the Cabibbo-Kobayashi-Maskawa (CKM) matrix element
$|V_{cb}|$~\cite{bclnu_vub} and the \bxsg branching
fraction~\cite{bsg_th} has reached the 2\% and 7\% levels,
respectively.  In order for any unambiguous statement to be made
regarding the presence of new physics, the inclusive experimental
measurements must be equally precise.

Although not sensitive to new physics contributions, the various
spectra from \bxclnu and \bxsg decays can be used to extract
fundamental SM parameters.  The OPE cannot be used in predicting the
spectral shapes, however, as the expansion breaks down at the phase
space endpoints, where the non-perturbative terms become significant.
To avoid this complication, integrated quantities (moments) instead of
the spectrum itself can be used to compare theory and experiment.  

% It has also been shown that the spectrum can be analytically expressed
% as a convolution of the perturbative contributions and a shape
% function, which incorporates all non-perturbative contributions.  The
% parameters of the shape function, which is not derivable from first
% principles, can be constrained by measurements of the \bxclnu and
% \bxsg spectra.

We will present a mini-review of inclusive measurements of \bxclnu and
\bxsg decays, primarily at the \PB-factories, and particularly
focusing on the \PcX mass moments, the \bxsg branching fraction, and
the \bxsdg direct \CP-asymmetry \acp, which is predicted to be nearly
zero in the SM~\cite{bdsg_th}.  We will also discuss the extraction of
heavy quark parameters from the moments measurements.

\section{Mass Moments from Semileptonic \bxclnu Decays}

The methods used in reconstructing \bxclnu decays at Belle and \babar\
are similar.  To suppress backgrounds from continuum events ($\Pep\Pem
\to \Pq\Paq$, where $\Pq = \Pqu,\Pqd,\Pqs,\Pqc$), one of the
\PB-mesons ($\PB_{reco}$) is reconstructed in fully hadronic final
states.  The large semileptonic branching fraction of the signal \PB
($\PB_{SL}$) makes the hadronic reconstruction method a statistically
feasible approach.  The signature of the semileptonic \PB decay is the
presence of a lepton with high energy, typically greater than 0.7 or
0.8~\gev in the \PB rest frame.  The remaining tracks and calorimeter
clusters in the event not used in tag reconstruction are combined to
form the final state \PcX hadronic system.

After the event selection, the remaining background can be classified
into three categories: combinatorial backgrounds, where $\PB_{reco}$
has been misreconstructed with particles from the $\PB_{SL}$;
continuum backgrounds; and residual backgrounds, where $\PB_{reco}$
was properly reconstructed, but $\PB_{SL}$ is reconstructed with
non-signal decays from misidentified leptons, cascade leptons, or
\bxulnu decays.

The Belle result uses $152 \times 10^6$ \BB events~\cite{belle_bclnu}
and removes continuum backgrounds by using data taken 60~\mev below
the \PgUc resonance (off-resonance data).  The signal and remaining
backgrounds are modeled with Monte Carlo (MC) samples.  Before the
\PcX mass moments are measured, detector resolution effects are
deconvoluted from the spectrum using a Singular Value Decomposition
(SVD) algorithm.  The moments and corresponding uncertainties are
calculated using the formulae
\begin{equation}
\left< M_X^k\right> = \frac{\sum_i (M_X^k)_i n_i}{\sum_i n_i}
\ \ \ \mbox{and} \ \ \ 
\sigma^2\left(\left< M_X^k\right>\right) = \frac{\sum_{i,j}(M_X^k)_i X_{ij}(M_X^k)_j}{\left(\sum_i n_i\right)^2}
\end{equation}
where $(M_X^k)_i$ is the central value of bin $i$ of the unfolded
spectrum, $n_i$ is the contribution to the spectrum at bin $i$, and
$X$ is the covariance matrix.  Belle measures moments corresponding to
$k=2, k=4$, and also the centralized mass moment $(M_X^2 -
\left<M_X^2\right>)^2$.

The \babar\ analysis is performed with $232 \times 10^6$ \BB
pairs~\cite{babar_bclnu}. Instead of using off-resonance data (as in
the Belle analysis), the \babar\ analysis uses a threshold function to
parameterize continuum as well as combinatorial background.  The
residual background and signal are modeled using MC samples.  Detector
resolution effects are taken into account through explicit
calibrations to the invariant \PcX mass.  Using MC samples, the
calibrations are parameterized using the form $M_{X,reco}^k = A +
B\times M_{X,calib}^k$, where $A$ and $B$ are constants that depend on
the energy imbalance in the event, the \PcX multiplicity, the minimum
lepton energy, and moment order $k$.  The masses $M_{X,reco}^k$ and
$M_{X,calib}^k$ correspond to the reconstructed, and calibrated mass
quantities, respectively.  A separate MC control sample of exclusive
semileptonic decays is used to validate the calibration procedure, the
results of which show good agreement between $M_{X,reco}^k$ and
$M_{X,calib}^k$.

A comparison of the results from both experiments for $k=2$ is shown
in Table~1.  The systematic errors are similar in both analyses.  They
arise from uncertainties in the assumption of the background
normalization, the variations of the \HepProcess{\PB \to
\PD^{(*)}\Pl\Pgn} branching fraction and form factors, and the
normalization of the other signal contributions, including
non-resonant final states.  Whereas Belle assigns an uncertainty due
to the unfolding parameter from the SVD algorithm, the \babar\
analysis takes into account the uncertainty of the calibration
procedure.  The \PcX moments, combined with the lepton energy moments
and the \bxsg photon energy moments, are used to extract the heavy
quark parameters (Section~\ref{sec:Extraction}).

\begin{table*}[!tb]  \label{massMoments}
 \begin{center}
  \begin{tabular}{ccc} \hline \hline
   $E_{\min}^* (\gev)$       & Belle Analysis $(\gev^2/c^4)$ &  \babar\ Analysis $(\gev^2/c^4)$    \\ \hline
   0.7 & $4.403 \ \pm \ 0.036 \ \pm \ 0.052$ & $\mbox{------------}$ \\  
   0.9 & $4.353 \ \pm \ 0.032 \ \pm \ 0.041$ & $4.416 \ \pm \ 0.027 \ \pm \ 0.063$ \\  
   1.1 & $4.293 \ \pm \ 0.028 \ \pm \ 0.029$ & $4.354 \ \pm \ 0.026 \ \pm \ 0.063$ \\  
   1.3 & $4.213 \ \pm \ 0.027 \ \pm \ 0.024$ & $4.281 \ \pm \ 0.027 \ \pm \ 0.061$ \\  
   1.5 & $4.144 \ \pm \ 0.028 \ \pm \ 0.022$ & $4.220 \ \pm \ 0.031 \ \pm \ 0.070$ \\  
   1.7 & $4.056 \ \pm \ 0.033 \ \pm \ 0.022$ & $4.158 \ \pm \ 0.040 \ \pm \ 0.094$ \\  
   1.9 & $3.996 \ \pm \ 0.041 \ \pm \ 0.021$ & $4.136 \ \pm \ 0.069 \ \pm \ 0.142$ \\  \hline
  \end{tabular}
  \caption{Second-order mass moments for various minimum lepton energy
  requirements ($E_{\min}^*$).  The \babar\ analysis measures moments
  in 100~\mev increments, starting at a minimum lepton energy of
  0.8~\gev--the other measurements are omitted for brevity.}
 \end{center}
\end{table*}

\section{Branching Fraction and \acp of \bxsg Decays}

The principal signature of an inclusive \bxsg decay is the high-energy
photon.  Two approaches have been used: a semi-inclusive method where
multiple \PsX final states are reconstructed to approximate an
inclusive process; and the fully-inclusive method where the \PsX is
not reconstructed. Although the semi-inclusive method results in small
backgrounds, assumptions must be made regarding the fraction of final
states that were not reconstructed, introducing an unavoidable model
dependence.  In contrast, the fully-inclusive method suffers from
large experimental backgrounds, but is sensitive to all hadronic final
states within the measured energy region.

The fully-inclusive analyses at \babar\ and Belle are
similar~\cite{belle_bsg}.  Both analyses impose a minimum photon
energy cut to remove significant backgrounds, principally from photon
daughters of \Pgpz mesons in continuum and $\Pep\Pem \to \Pgtp\Pgtm$
events, and non-signal \PB decays.  To remove photons from \Pgpz and
\Pgh decays, the invariant mass of the photon candidate and any other
photon in the event is calculated; the event is rejected if the
invariant mass is consistent with the nominal \Pgpz or \Pgh mass.
Photons from continuum events are suppressed by making requirements on
high-momentum leptons (lepton tagging), which are unlikely to come
from continuum events, and by exploiting the different event
topologies between continuum and \BB events by using multivariate
algorithms.
%
% \begin{itemize}
% \item Leptons from continuum events tend to peak at lower momenta than
% those from semileptonic \PB decays.  By imposing a minimum lepton
% momentum cut, a significant fraction of continuum events are removed
% ($\sim 98\%$), yet $\sim 10\%$ of the signal remains.
% \item The small boost of the \PB mesons in the \PgUc rest frame
% translates to nearly isotropic distributions of \PB decays.  In
% contrast, the additional kinetic energy in continuum events gives rise
% to jet-like final states in opposite hemispheres of the detector.  The
% event topology can therefore be used in multivariate algorithms to
% discriminate between signal-like and background-like events.
% \end{itemize}

% Whereas \babar\ exclusively uses lepton tagging, Belle performs a
% hybrid analysis with lepton-tagged and non-lepton-tagged events, which
% are then combined to give an overall result.  The event topology
% variables (along with the lepton tag variables) are included in a
% neural network in the \babar\ analysis to discriminate between signal
% and background (primarily continuum); Belle uses a simple linear
% discriminant to account for differences in event shapes between signal
% and background.
% 
To remove the remaining background, continuum events are subtracted
from the photon energy spectrum using luminosity-scaled off-resonance
data---data taken 60 (40)~\mev below the \PgUc resonance at Belle
(\babar). Backgrounds from \PB decays are removed using data-corrected
MC simulations.  The contribution from \bxdg decays must be removed
from the resulting energy spectrum, and various procedures must be
employed to remove effects from the boost of the \PB to the \PgUc rest
frame, and from the detector resolution, which is convoluted with the
true photon energy spectrum.

\begin{table*}[t]
 \begin{center}
  \label{tab:BBcomposition}
  \begin{tabular}{|l|c|c|c|} \hline 
   Background        & Belle Analysis &  \babar\ Analysis       \\ \cline{2-3}
   Process           & $1.7 < \egcms < 2.8 \gev$ & $1.8 < \egcms < 2.8 \gev$ \\ \hline
   \HepProcess{\PB \to X\Pgpz }       & $0.597^*$                 & $0.613^*$              \\
   \HepProcess{\PB \to X\Pgh  }       & $0.199^*$                 & $0.192^*$              \\  \cline{2-2}
   \HepProcess{\PB \to X\Pgo  }       & \multirow{2}{*}{ $\uparrow$ }                     & $0.027^*$              \\
   \HepProcess{\PB \to X\Pghpr}       &                     & $0.008^*$              \\
   \HepProcess{\PB \to X\PJgy }       & $0.111$                  & $0.007$               \\
   \HepProcess{\PB \to X\Pepm(\Pgg) } & \multirow{2}{*}{$\downarrow$}                      & $0.062$               \\
   Final State Radiation (FSR)        &                       & $0.019$               \\\hline
   Fake Photon: \Pepm                 & $0.041$                  & $0.033$               \\
   Fake Photon: \PKzL and \Pan        & $0.020$                  & $0.025^*$       \\ \hline
   Other                              & $0.032$                  & $0.014$               \\ \hline
  \end{tabular}
  \caption{The \BB background composition according to Monte Carlo
   simulation after all selection cuts in the signal regions for the
   Belle and \babar\ analyses.  The variable $\egcms$ is the candidate
   photon energy in the \PgUc rest frame.  Fractions followed by an
   asterisk represent backgrounds that are corrected using an
   appropriate data control sample.  The backgrounds from
   \HepProcess{\PB \to X\Pgo } to FSR are grouped together in the
   Belle analysis and comprise 11.1\% of the total \BB background.
   Note that the antineutron and not the \PKzL background component is
   corrected in the \babar\ analysis.}
 \end{center}
\end{table*}

The breakdown of \PB backgrounds in the \babar\ and Belle analyses is
shown in Table~2.  The fractions are given in the signal regions of
the respective analyses.  The majority of \BB background photons after
the event selection are from \pizeta decays where the photon partner
was not reconstructed, allowing the photon candidate to slip past the
\pizeta vetoes.  Belle corrects for photon backgrounds just from \Pgpz
and \Pgh decays using an uncorrelated data sample of inclusive \pizeta
decays.  In addition to using dedicated control samples for corrected
\pizeta backgrounds, \babar\ also corrects for photons that arise from
\Pgo and \Pghpr decays, as well as for false photon signatures by
electrons and antineutrons.

The efficiency-corrected and unfolded photon energy spectrum as
measured by Belle is shown in the left plot of Figure~1.  The
branching fraction obtained is $\mathcal{B}(\bxsg) = (3.45 \pm
0.15_{stat.} \pm 0.40_{syst.})\times 10^{-4}$ for a minimum photon
energy of 1.7~\gev in the \PB rest frame.  Reference~\cite{belle_bsg}
quotes branching fraction results for various minimum photon energy
requirements, as well as the corresponding first and
second-centralized energy moments.  The dominant systematic errors
arise from the uncertainty on the \BB background estimation and
data-based corrections.  Additional systematic uncertainties enter
from the unfolding algorithm, the photon detection efficiency, the
removal of \bxdg contamination, and the transformation from the \PgUc
to the \PB rest frame.

The preliminary photon energy spectrum measured at \babar\ is shown in
the right plot of Figure~1.  The energy region above 2.9\gev is used
to validate the off-resonance subtraction procedure.  The various \BB
MC corrections are validated in the energy region $1.53 < \egcms < 1.8
\gev$, which is composed almost entirely of \BB background after
continuum background subtraction.  The flavor of the signal \Pqb-quark
is identified by the charge of the tag lepton: $\Plp(\Plm) \implies
\Pqb(\Paqb)$.  The photon energy spectrum is then divided according to
the lepton charge.  The \acp is then:
\begin{equation}
\acp(\bxsdg) = \frac{1}{1-2\omega}\frac{N^+ - N^-}{N^+ + N^-}
\end{equation}
where $\omega$ accounts for dilution effects, and $N^{+(-)}$ are the
events tagged with an $\Plp(\Plm)$.  

\begin{figure}[t]\label{bsg_spectra}
   \includegraphics[width=0.425\textwidth]{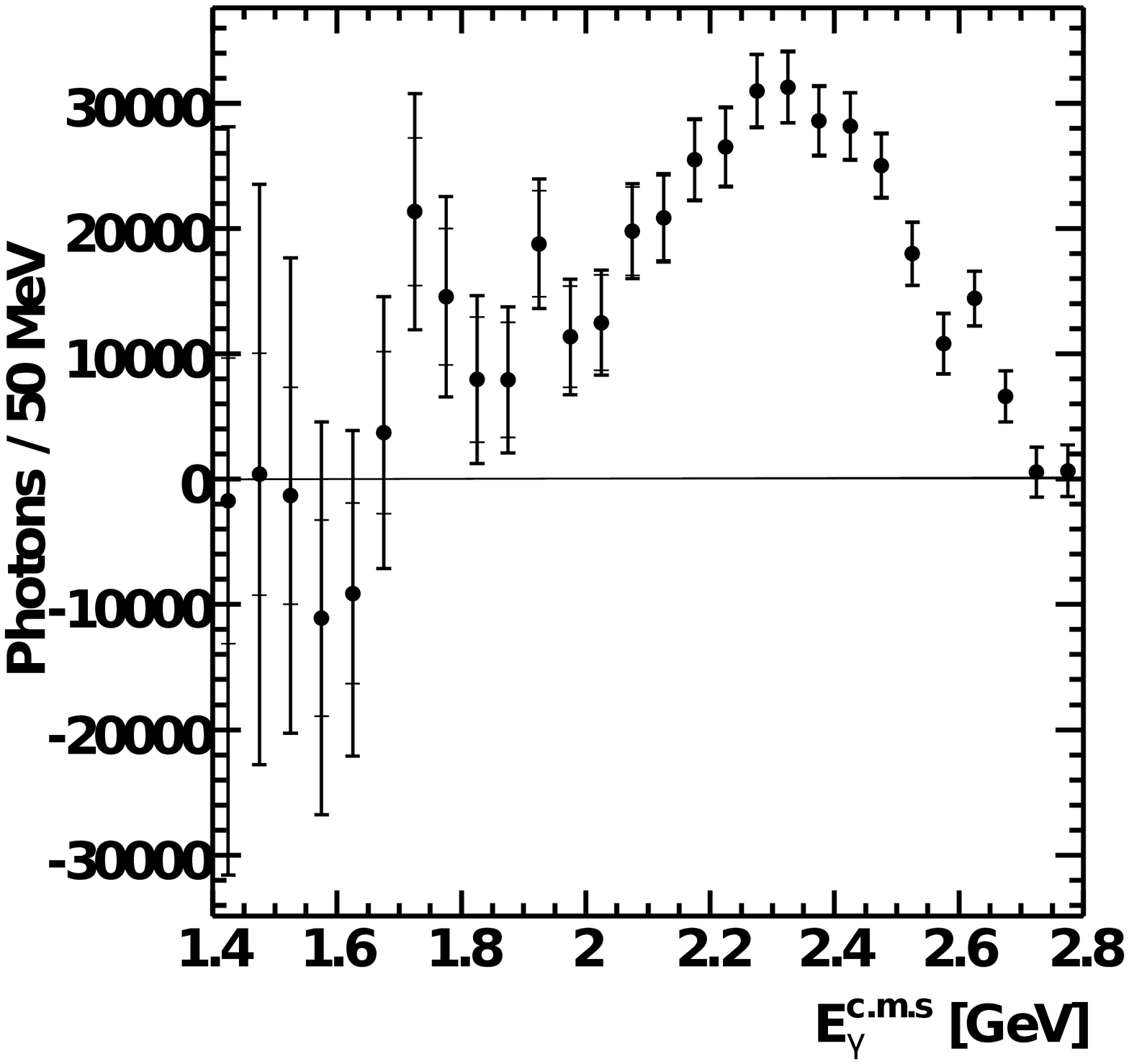}
   \includegraphics[width=0.575\textwidth]{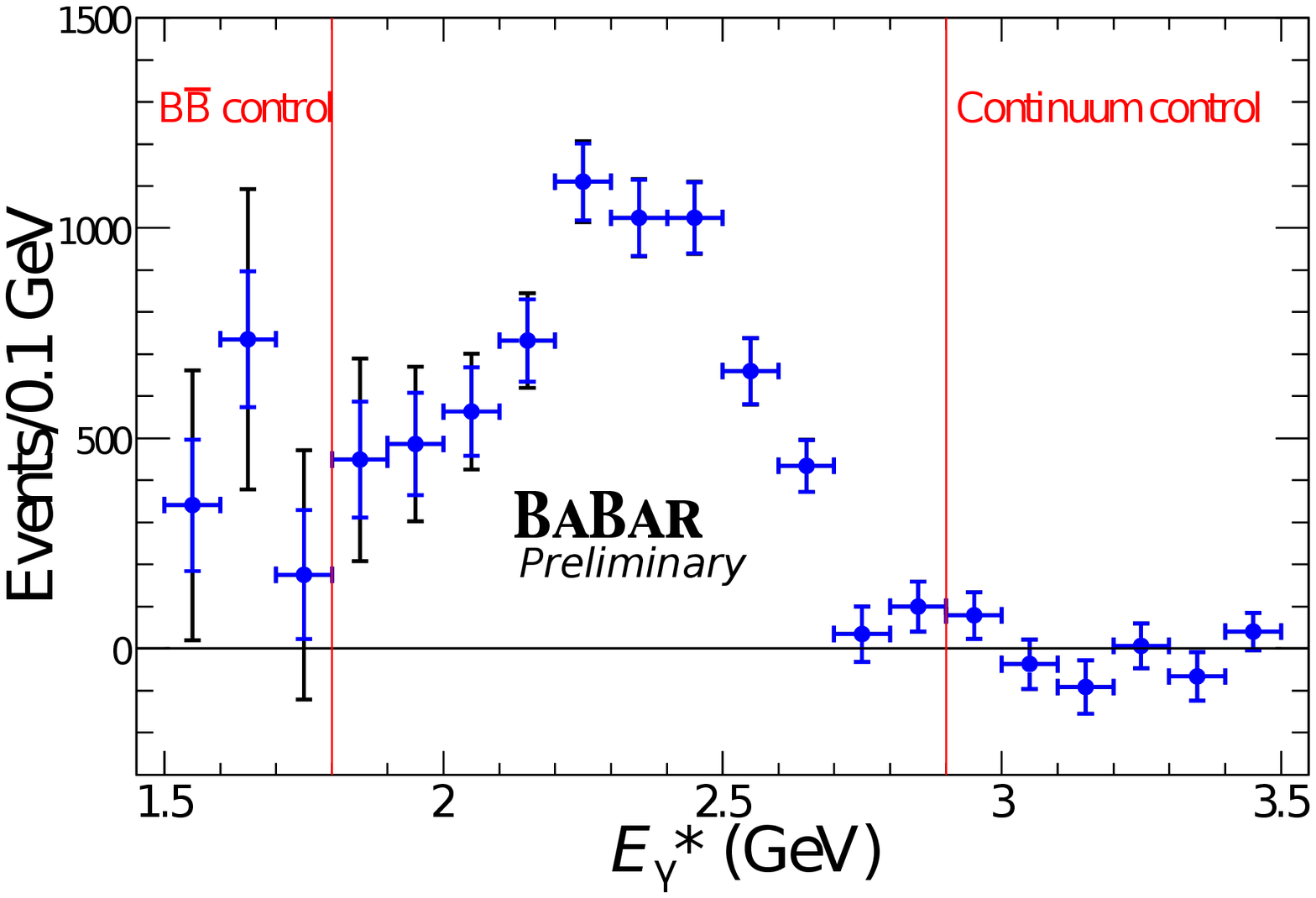}
   \caption{{\bf(left)} Measured photon energy spectrum from Belle,
   corrected for efficiency and detector effects; {\bf(right)}
   preliminary photon spectrum at \babar\ after all selection cuts.  Inner error bars, where visible,
    represent the statistical contribution to the total error (outer
    error bars).}
\end{figure}

To reduce the sensitivity to the systematic uncertainties of the \BB
background, the \acp is extracted with a photon energy cut of $\egcms
> 2.1 \gev$.  The dominant uncertainty is therefore due to the limited
statistics of the off-resonance data subtraction.  The measured yields
are $N^{+(-)} = 2397 \pm 151_{stat.} (2623 \pm 158_{stat.})$, giving
rise to a raw \acp of $0.045 \pm 0.044_{stat.}$.  The dilution term
$\omega = 0.131 \pm 0.0064_{syst.}$ arises from wrong-sign leptons,
which result from \PBz-\PaBz oscillations, $\PB \to \PD \to X\Pl\nu$
cascade decays, and lepton misidentification.  
% The systematic errors come in two types:
%
% \begin{itemize}
% \item {\bf Multiplicative:} errors that are proportional to the
% central value of the raw \acp.  Such errors include uncertainties on
% the dilution term $\omega$ and the overall \BB background yield.
% \item {\bf Additive:} errors that constitute a direct bias on the
% measured \acp.  These errors result from \CP asymmetry in the \BB
% background, as well as a charge asymmetry in the lepton
% reconstruction.
% \end{itemize}
%
Accounting for these effects and also additional potential biases from
the \BB subtraction, the preliminary result at \babar\ is
$\acp(\bxsdg) = 0.056 \pm 0.063$, consistent with SM expectation.  A
comparison to previous measurements is shown in Figure~2.

\begin{figure}[t]\label{acpComp}
  \begin{center}
    \includegraphics[width=0.8\textwidth]{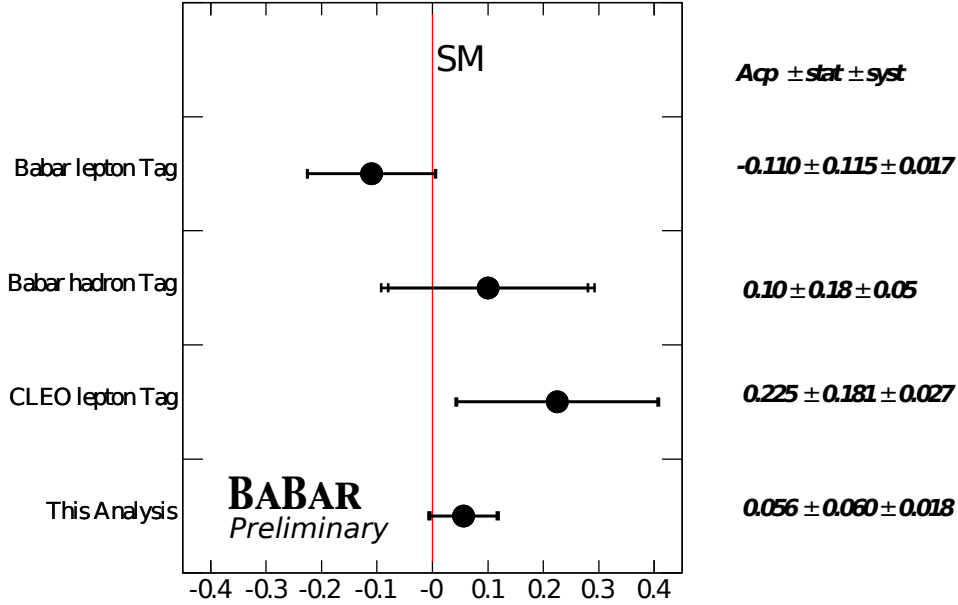}
    \caption{Comparison of previous $\acp(\bxsdg)$ measurements to the
    preliminary \babar\ result.  Inner error bars, where visible,
    represent the statistical contribution to the total error (outer
    error bars).}
  \end{center}
\end{figure}

\section{Extraction of Heavy Quark Parameters} 
\label{sec:Extraction}

Due to its heavy mass, the \Pqb-quark field can be expanded
non-relativistically.  In the context of the Heavy Quark Expansion
(HQE), various parameters arise, which characterize the motion of the
\Pqb-quark inside the \PB meson.  In the kinetic scheme, at the lowest
orders appear expectation values of dimension-five and -six operators:
$\mu_\pi^2$ (Fermi motion), $\mu_G^2$ (\PB - \PBst splitting),
$\rho_{LS}^3$ (spin-orbit coupling) and $\rho_{D}^3$ (Darwin term).

To extract the values of these parameters, along with the value of the
\Pqb-quark mass \mb and the CKM matrix element $|V_{cb}|$, the \PcX
mass moments and lepton energy moments from \bxclnu decays and the
photon energy moments from \bxsg decays are combined into a fit, based
on a $\chi^2$ minimization technique~\cite{hqe_fit}.  A vector of
experimental moments ($\mathbf{M}_\mathrm{exp}$) is compared with the
analytic predictions from the HQE ($\mathbf{M}_\mathrm{HQE}$).  The
constructed $\chi^2$:
\begin{equation}
\chi^2 = \left(\mathbf{M}_\mathrm{exp}-\mathbf{M}_\mathrm{HQE}\right)^T C_\mathrm{tot}^{-1}\left(\mathbf{M}_\mathrm{exp}-\mathbf{M}_\mathrm{HQE}\right)
\end{equation}
is minimized to obtain the HQE parameters.  The sum of the
experimental and theoretical covariance matrices is represented by
$C_\mathrm{tot}$.

The fit results from measurements of the individual Belle and \babar\
experiments are presented in
References~\cite{babar_bclnu,belle_moments}.  The $|V_{cb}|$ \vs $\mb$
and $\mu_\pi^2$ \vs $\mb$ results from a global fit by the Heavy
Flavor Averaging Group (HFAG) are shown in Figure~3~\cite{hfag}.  The
global fit includes measurements from various experiments, but
excludes the most recent photon energy moments from Belle.  Inclusion
of the photon energy moments gives rise to a roughly $1\sigma$ tension
between the results with and without the photon moments.  This tension
is a source of much discussion among the theory and experimental
communities, and further discussion is beyond the scope of this
Proceedings contribution.

\begin{figure}[hbt]
  \includegraphics[width=0.48\textwidth]{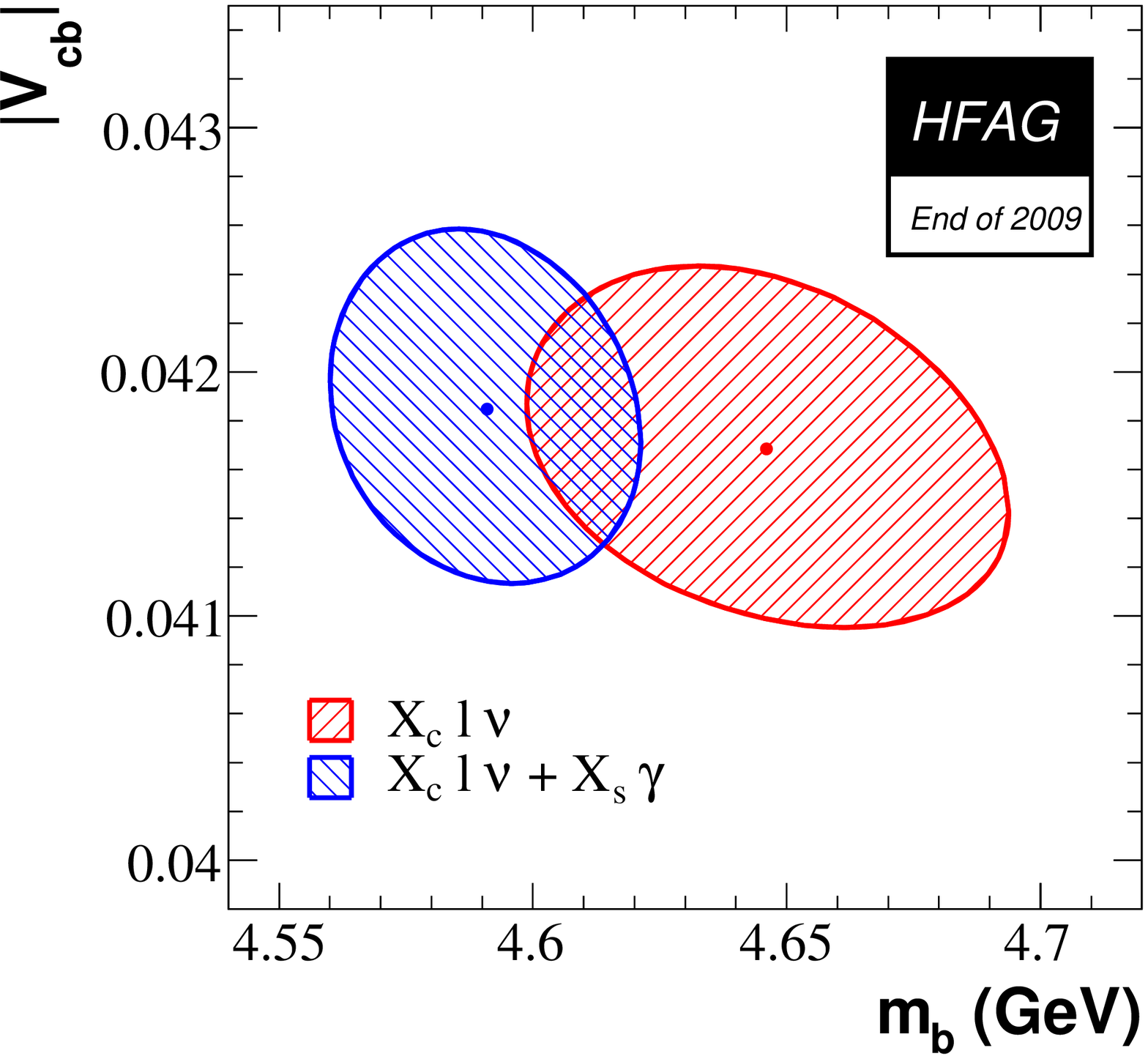}
  \includegraphics[width=0.48\textwidth]{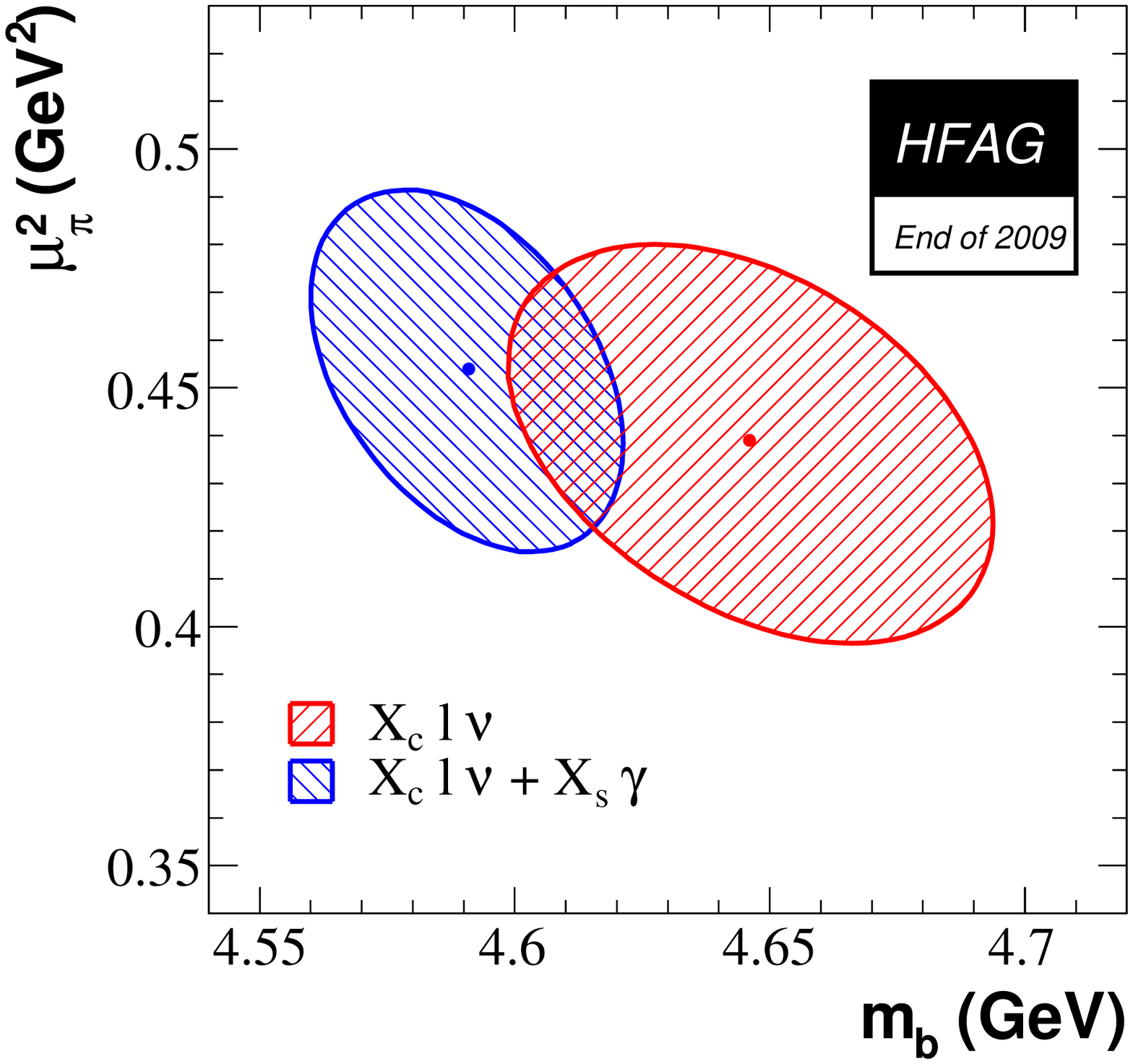}
  \caption{ Global fit results by HFAG.  Shown are the $\Delta\chi^2 =
  1$ contours for the fit with just the semileptonic moments (labeled
  $\PcX\Pl\Pgn$) and the fit that includes the photon energy moments
  (labeled $\PcX\Pl\Pgn + \PsX\gamma$).}
\end{figure}

\section{Summary \& Acknowledgments}

We have argued that inclusive \bxclnu and \bxsg decays are probes of
the Standard Model in extracting fundamental SM parameters ($|V_{cb}|$
and \mb) and HQE parameters ($\mu_\pi^2$, $\mu_G^2$, $\rho_{LS}^3$ and
$\rho_{D}^3$) through fits to the measured moments.  Radiative \bxsg
decays are also sensitive to new physics by new physics particles
propagating in the penguin diagram.  The \bxsg branching fraction and
direct \CP-asymmetry results presented, however, are consistent with
SM expectations.

The author would like to thank the organizers of the Heavy Quarks and
Leptons conference for an enjoyable experience, and also Vittorio
Lubicz for explaining the status of the lattice determination of the
hadronic form factor in \bkg decays.  He also thanks Kevin Flood for
helpful discussions, and appreciates the support from Colin Jessop.

\end{document}